# 1. NEUTRINO MASS, MIXING,
# AND FLAVOR CHANGE*

Revised March 2008 by B. Kayser[†]
Fermilab, P.O. Box 500, Batavia, IL 60510, USA

There is now compelling evidence that atmospheric, solar, accelerator, and reactor neutrinos change from one flavor to another. This implies that neutrinos have masses and that leptons mix. In this review, we discuss the physics of flavor change and the evidence for it, summarize what has been learned so far about neutrino masses and leptonic mixing, consider the relation between neutrinos and their antiparticles, and discuss the open questions about neutrinos to be answered by future experiments.

***I. The physics of flavor change:*** If neutrinos have masses, then there is a spectrum of three or more neutrino mass eigenstates, $\nu_1$, $\nu_2$, $\nu_3$, ..., that are the analogues of the charged-lepton mass eigenstates, $e$, $\mu$, and $\tau$. If leptons mix, the weak interaction coupling the $W$ boson to a charged lepton and a neutrino can couple any charged-lepton mass eigenstate $\ell_\alpha$ to any neutrino mass eigenstate $\nu_i$. Here, $\alpha = e, \mu$, or $\tau$, and $\ell_e$ is the electron, *etc.* The amplitude for the decay of a real or virtual $W^+$ to yield the specific combination $\ell_\alpha^+ + \nu_i$ is $U_{\alpha i}^*$, where $U$ is the unitary leptonic mixing matrix [1]. Thus, the neutrino state created in the decay $W^+ \to \ell_\alpha^+ + \nu$ is the state

$$|\nu_\alpha\rangle = \sum_i U_{\alpha i}^* |\nu_i\rangle \ . \tag{1.1}$$

This superposition of neutrino mass eigenstates, produced in association with the charged lepton of "flavor" $\alpha$, is the state we refer to as the neutrino of flavor $\alpha$. Assuming $CPT$ invariance, the unitarity of $U$ guarantees that the only charged lepton a $\nu_\alpha$ can create in a detector is an $\ell_\alpha$, with the same flavor as the neutrino. Eq. (1.1) may be inverted to give

$$|\nu_i\rangle = \sum_\beta U_{\beta i} |\nu_\beta\rangle \ , \tag{1.2}$$

which expresses the mass eigenstate $\nu_i$ as a superposition of the neutrinos of definite flavor.

While there are only three (known) charged lepton mass eigenstates, it may be that there are more than three neutrino mass eigenstates. If, for example, there are four $\nu_i$, then one linear combination of them,

$$|\nu_s\rangle = \sum_i U_{si}^* |\nu_i\rangle \ , \tag{1.3}$$

does not have a charged-lepton partner, and consequently does not couple to the Standard Model $W$ boson. Indeed, since the decays $Z \to \nu_\alpha \overline{\nu}_\alpha$ of the Standard Model $Z$ boson have been found to yield only three distinct neutrinos $\nu_\alpha$ of definite flavor [2], $\nu_s$ does not couple to the $Z$ boson either. Such a neutrino, which does not have any Standard Model weak couplings, is referred to as a "sterile" neutrino.

Neutrino flavor change is the process $\nu_\alpha \to \nu_\beta$, in which a neutrino born with flavor $\alpha$ becomes one of a different flavor $\beta$ while propagating in vacuum or in matter. This process, often referred to as neutrino oscillation, is quantum mechanical to its core.

Rather than present a full wave packet treatment [3], we shall give a simpler description that captures all the essential physics. We begin with oscillation in vacuum, and work in the neutrino mass eigenstate basis. Then the neutrino that travels from the source to the detector is one or another of the mass eigenstates $\nu_i$. The amplitude for the oscillation $\nu_\alpha \to \nu_\beta$, Amp $(\nu_\alpha \to \nu_\beta)$, is a coherent sum over the contributions of all the $\nu_i$, given by

$$\text{Amp}\left(\nu_\alpha \to \nu_\beta\right) = \sum_i U_{\alpha i}^* \, \text{Prop}\,(\nu_i)\, U_{\beta i} \quad . \tag{1.4}$$

In the contribution $U_{\alpha i}^* \,\text{Prop}\,(\nu_i)\, U_{\beta i}$ of $\nu_i$ to this sum, the factor $U_{\alpha i}^*$ is the amplitude for the neutrino $\nu_\alpha$ to be the mass eigenstate $\nu_i$ [see Eq. (1.1)], the factor $\text{Prop}\,(\nu_i)$ is the amplitude for this $\nu_i$ to propagate from the source to the detector, and the factor $U_{\beta i}$ is the amplitude for the $\nu_i$ to be a $\nu_\beta$ [see Eq. (1.2)]. From elementary quantum mechanics, the propagation amplitude $\text{Prop}\,(\nu_i)$ is $\exp[-im_i\tau_i]$, where $m_i$ is the mass of $\nu_i$, and $\tau_i$ is the proper time that elapses in the $\nu_i$ rest frame during its propagation. By Lorentz invariance, $m_i\tau_i = E_i t - p_i L$, where $L$ is the lab-frame distance between the neutrino source and the detector, $t$ is the lab-frame time taken for the beam to traverse this distance, and $E_i$ and $p_i$ are, respectively, the lab-frame energy and momentum of the $\nu_i$ component of the beam.

In the probability $P(\nu_\alpha \to \nu_\beta) = |\text{Amp}\,(\nu_\alpha \to \nu_\beta)|^2$ for the oscillation $\nu_\alpha \to \nu_\beta$, only the *relative* phases of the propagation amplitudes $\text{Prop}\,(\nu_i)$ for different mass eigenstates will have physical consequences. From the discussion above, the relative phase of $\text{Prop}\,(\nu_i)$ and $\text{Prop}\,(\nu_j)$, $\delta\phi_{ij}$, is given by

$$\delta\phi_{ij} = \left(p_i - p_j\right)L - \left(E_i - E_j\right)t \quad . \tag{1.5}$$

In practice, experiments do not measure the transit time $t$. However, Lipkin has shown [4] that, to an excellent approximation, the $t$ in Eq. (1.5) may be taken to be $L/\bar{v}$, where

$$\bar{v} = \frac{p_i + p_j}{E_i + E_j} \tag{1.6}$$

is an approximation to the average of the velocities of the $\nu_i$ and $\nu_j$ components of the beam. Then

$$\delta\phi_{ij} \cong \frac{p_i^2 - p_j^2}{p_i + p_j}L - \frac{E_i^2 - E_j^2}{p_i + p_j}L \cong \left(m_j^2 - m_i^2\right)\frac{L}{2E} \quad , \tag{1.7}$$

where, in the last step, we have used the fact that for highly relativistic neutrinos, $p_i$ and $p_j$ are both approximately equal to the beam energy $E$. We conclude that all the relative phases in Amp $(\nu_\alpha \to \nu_\beta)$, Eq. (1.4), will be correct if we take $\text{Prop}\,(\nu_i) = \exp\left(-im_i^2 L/2E\right)$, so that

$$\text{Amp}\left(\nu_\alpha \to \nu_\beta\right) = \sum_i U_{\alpha i}^* \, e^{-im_i^2 L/2E}\, U_{\beta i} \quad . \tag{1.8}$$

Squaring, and making judicious use of the unitarity of $U$, we then find that

$$
\begin{aligned}
P\left(\nu_\alpha \rightarrow \nu_\beta\right) = {} & \delta_{\alpha\beta} \\
& -4 \sum_{i>j} \Re \left(U_{\alpha i}^* U_{\beta i} U_{\alpha j} U_{\beta j}^*\right) \sin^2 \left[1.27 \, \Delta m_{ij}^2 \left(L/E\right)\right] \\
& +2 \sum_{i>j} \Im \left(U_{\alpha i}^* U_{\beta i} U_{\alpha j} U_{\beta j}^*\right) \sin \left[2.54 \, \Delta m_{ij}^2 \left(L/E\right)\right] \ .
\end{aligned}
\tag{1.9}
$$

Here, $\Delta m_{ij}^2 \equiv m_i^2 - m_j^2$ is in $\mathrm{eV}^2$, $L$ is in km, and $E$ is in GeV. We have used the fact that when the previously omitted factors of $\hbar$ and $c$ are included,

$$
\Delta m_{ij}^2 \left(L/4E\right) \simeq 1.27 \, \Delta m_{ij}^2 \left(\mathrm{eV}^2\right) \frac{L\left(\mathrm{km}\right)}{E\left(\mathrm{GeV}\right)} \ .
\tag{1.10}
$$

Assuming that $CPT$ invariance holds,

$$
P\left(\overline{\nu}_\alpha \rightarrow \overline{\nu}_\beta\right) = P\left(\nu_\beta \rightarrow \nu_\alpha\right) \ .
\tag{1.11}
$$

But, from Eq. (1.9) we see that

$$
P\left(\nu_\beta \rightarrow \nu_\alpha; U\right) = P\left(\nu_\alpha \rightarrow \nu_\beta; U^*\right) \ .
\tag{1.12}
$$

Thus, when $CPT$ holds,

$$
P\left(\overline{\nu}_\alpha \rightarrow \overline{\nu}_\beta; U\right) = P\left(\nu_\alpha \rightarrow \nu_\beta; U^*\right) \ .
\tag{1.13}
$$

That is, the probability for oscillation of an antineutrino is the same as that for a neutrino, except that the mixing matrix $U$ is replaced by its complex conjugate. Thus, if $U$ is not real, the neutrino and antineutrino oscillation probabilities can differ by having opposite values of the last term in Eq. (1.9). When $CPT$ holds, any difference between these probabilities indicates a violation of $CP$ invariance.

As we shall see, the squared-mass splittings $\Delta m_{ij}^2$ called for by the various reported signals of oscillation are quite different from one another. It may be that one splitting, $\Delta M^2$, is much bigger than all the others. If that is the case, then for an oscillation experiment with $L/E$ such that $\Delta M^2 L/E = \mathcal{O}(1)$, Eq. (1.9) simplifies considerably, becoming

$$
P\left(\overset{\scriptscriptstyle(-)}{\nu}_\alpha \rightarrow \overset{\scriptscriptstyle(-)}{\nu}_\beta\right) \simeq S_{\alpha\beta} \sin^2 \left[1.27 \, \Delta M^2 \left(L/E\right)\right]
\tag{1.14}
$$

for $\beta \neq \alpha$, and

$$
P\left(\overset{\scriptscriptstyle(-)}{\nu}_\alpha \rightarrow \overset{\scriptscriptstyle(-)}{\nu}_\alpha\right) \simeq 1 - 4 \, T_\alpha \left(1 - T_\alpha\right) \sin^2 \left[1.27 \, \Delta M^2 \left(L/E\right)\right] \ .
\tag{1.15}
$$

Here,

$$
S_{\alpha\beta} \equiv 4 \left| \sum_{i \ \mathrm{Up}} U_{\alpha i}^* U_{\beta i} \right|^2
\tag{1.16}
$$

and

$$T_\alpha \equiv \sum_{i \ \mathrm{Up}} |U_{\alpha i}|^2 \ , \tag{1.17}$$

where "$i$ Up" denotes a sum over only those neutrino mass eigenstates that lie *above* $\Delta M^2$ or, alternatively, only those that lie *below* it. The unitarity of $U$ guarantees that summing over either of these two clusters will yield the same results for $S_{\alpha\beta}$ and for $T_\alpha(1 - T_\alpha)$.

The situation described by Eqs. (1.14)–(1.17) may be called "quasi-two–neutrino oscillation." It has also been called "one mass scale dominance" [5]. It corresponds to an experiment whose $L/E$ is such that the experiment can "see" only the big splitting $\Delta M^2$. To this experiment, all the neutrinos above $\Delta M^2$ appear to be a single neutrino, as do all those below $\Delta M^2$.

The relations of Eqs. (1.14)–(1.17) apply to a three-neutrino spectrum in which one of the two squared-mass splittings is much bigger than the other one. If we denote by $\nu_3$ the neutrino that is by itself at one end of the large splitting $\Delta M^2$, then $S_{\alpha\beta} = 4|U_{\alpha 3}U_{\beta 3}|^2$ and $T_\alpha = |U_{\alpha 3}|^2$. Thus, oscillation experiments with $\Delta M^2 L/E = \mathcal{O}(1)$ can determine the flavor fractions $|U_{\alpha 3}|^2$ of $\nu_3$.

The relations of Eqs. (1.14)–(1.17) also apply to the special case where, to a good approximation, only two mass eigenstates, and two corresponding flavor eigenstates (or two linear combinations of flavor eigenstates), are relevant. One encounters this case when, for example, only two mass eigenstates couple significantly to the charged lepton with which the neutrino being studied is produced. When only two mass eigenstates count, there is only a single splitting, $\Delta m^2$, and, omitting irrelevant phase factors, the unitary mixing matrix $U$ takes the form

$$U = \begin{array}{c} \\ \nu_\alpha \\ \nu_\beta \end{array} \begin{array}{cc} \nu_1 & \nu_2 \\ \left[ \begin{array}{cc} \cos\theta & \sin\theta \\ -\sin\theta & \cos\theta \end{array} \right] \end{array} \ . \tag{1.18}$$

Here, the symbols above and to the left of the matrix label the columns and rows, and $\theta$ is referred to as the mixing angle. From Eqs. (1.16) and (1.17), we now have $S_{\alpha\beta} = \sin^2 2\theta$ and $4\,T_\alpha(1 - T_\alpha) = \sin^2 2\theta$, so that Eqs. (1.14) and (1.15) become, respectively,

$$P\left(\overset{(-)}{\nu_\alpha} \to \overset{(-)}{\nu_\beta}\right) = \sin^2 2\theta \, \sin^2 \left[1.27 \, \Delta m^2 \, (L/E)\right] \tag{1.19}$$

with $\beta \neq \alpha$, and

$$P\left(\overset{(-)}{\nu_\alpha} \to \overset{(-)}{\nu_\alpha}\right) = 1 - \sin^2 2\theta \, \sin^2 \left[1.27 \, \Delta m^2 \, (L/E)\right] \ . \tag{1.20}$$

Many experiments have been analyzed using these two expressions. Some of these experiments actually have been concerned with quasi-two-neutrino oscillation, rather than a genuinely two-neutrino situation. For these experiments, "$\sin^2 2\theta$" and "$\Delta m^2$" have the significance that follows from Eqs. (1.14)–(1.17).

When neutrinos travel through matter (*e.g.*, in the Sun, Earth, or a supernova), their coherent forward-scattering from particles they encounter along the way can significantly modify their propagation [6]. As a result, the probability for changing flavor can be rather different than it is in vacuum [7]. Flavor change that occurs in matter, and that grows out of the interplay between flavor-nonchanging neutrino-matter interactions and neutrino mass and mixing, is known as the Mikheyev-Smirnov-Wolfenstein (MSW) effect.

To a good approximation, one can describe neutrino propagation through matter via a Schrödinger-like equation. This equation governs the evolution of a neutrino state vector with several components, one for each flavor. The effective Hamiltonian in the equation, a matrix $\mathcal{H}$ in neutrino flavor space, differs from its vacuum counterpart by the addition of interaction energies arising from the coherent forward neutrino-scattering. For example, the $\nu_e$–$\nu_e$ element of $\mathcal{H}$ includes the interaction energy

$$V = \sqrt{2}\, G_F N_e \ , \tag{1.21}$$

arising from $W$-exchange-induced $\nu_e$ forward-scattering from ambient electrons. Here, $G_F$ is the Fermi constant, and $N_e$ is the number of electrons per unit volume. In addition, the $\nu_e$–$\nu_e$, $\nu_\mu$–$\nu_\mu$, and $\nu_\tau$–$\nu_\tau$ elements of $\mathcal{H}$ all contain a common interaction energy growing out of $Z$-exchange-induced forward-scattering. However, when one is not considering the possibility of transitions to sterile neutrino flavors, this common interaction energy merely adds to $\mathcal{H}$ a multiple of the identity matrix, and such an addition has no effect on flavor transitions.

The effect of matter is illustrated by the propagation of solar neutrinos through solar matter. When combined with information on atmospheric neutrino oscillation, the experimental bounds on short-distance ($L \lesssim 1$ km) oscillation of reactor $\overline{\nu}_e$ [8] tell us that, if there are no sterile neutrinos, then only two neutrino mass eigenstates, $\nu_1$ and $\nu_2$, are significantly involved in the evolution of the solar neutrinos. Correspondingly, only two flavors are involved: the $\nu_e$ flavor with which every solar neutrino is born, and the effective flavor $\nu_x$ — some linear combination of $\nu_\mu$ and $\nu_\tau$ — which it may become. The Hamiltonian $\mathcal{H}$ is then a $2 \times 2$ matrix in $\nu_e$–$\nu_x$ space. Apart from an irrelevant multiple of the identity, for a distance $r$ from the center of the Sun, $\mathcal{H}$ is given by

$$\begin{aligned}
\mathcal{H} &= \mathcal{H}_V + \mathcal{H}_M\,(r) \\
&= \frac{\Delta m_\odot^2}{4E}
\begin{bmatrix} -\cos 2\theta_\odot & \sin 2\theta_\odot \\ \sin 2\theta_\odot & \cos 2\theta_\odot \end{bmatrix}
+ \begin{bmatrix} V(r) & 0 \\ 0 & 0 \end{bmatrix} \ .
\end{aligned} \tag{1.22}$$

Here, the first matrix $\mathcal{H}_V$ is the Hamiltonian in vacuum, and the second matrix $\mathcal{H}_M(r)$ is the modification due to matter. In $\mathcal{H}_V$, $\theta_\odot$ is the solar mixing angle defined by the two-neutrino mixing matrix of Eq. (1.18) with $\theta = \theta_\odot, \nu_\alpha = \nu_e$, and $\nu_\beta = \nu_x$. The splitting $\Delta m_\odot^2$ is $m_2^2 - m_1^2$, and for the present purpose we *define* $\nu_2$ to be the heavier of the two mass eigenstates, so that $\Delta m_\odot^2$ is positive. In $\mathcal{H}_M(r)$, $V(r)$ is the interaction energy of Eq. (1.21) with the electron density $N_e(r)$ evaluated at distance $r$ from the Sun's center.

From Eqs. (1.19–1.20) (with $\theta = \theta_\odot$), we see that two-neutrino oscillation in vacuum cannot distinguish between a mixing angle $\theta_\odot$ and an angle $\theta_\odot' = \pi/2 - \theta_\odot$. But these two mixing angles represent physically different situations. Suppose, for example, that $\theta_\odot < \pi/4$. Then, from Eq. (1.18) we see that if the mixing angle is $\theta_\odot$, the lighter mass eigenstate (defined to be $\nu_1$) is more $\nu_e$ than $\nu_x$, while if it is $\theta'_\odot$, then this mass eigenstate is more $\nu_x$ than $\nu_e$. While oscillation in vacuum cannot discriminate between these two possibilities, neutrino propagation through solar matter can do so. The neutrino interaction energy $V$ of Eq. (1.21) is of definite, positive sign [9]. Thus, the $\nu_e$–$\nu_e$ element of the solar $\mathcal{H}$, $-(\Delta m^2_\odot/4E)\cos 2\theta_\odot + V(r)$, has a different size when the mixing angle is $\theta'_\odot = \pi/2 - \theta_\odot$ than it does when this angle is $\theta_\odot$. As a result, the flavor content of the neutrinos coming from the Sun can be different in the two cases [10].

Solar and long-baseline reactor neutrino data establish that the behavior of solar neutrinos is governed by a Large-Mixing-Angle (LMA) MSW effect (see Sec. II). Let us estimate the probability $P(\nu_e \to \nu_e)$ that a solar neutrino that undergoes the LMA-MSW effect in the Sun still has its original $\nu_e$ flavor when it arrives at the Earth. We focus on the neutrinos produced by $^8$B decay, which are at the high-energy end of the solar neutrino spectrum. At $r \simeq 0$, where the solar neutrinos are created, the electron density $N_e \simeq 6 \times 10^{25}/\text{cm}^3$ [11] yields for the interaction energy $V$ of Eq. (1.21) the value $0.75 \times 10^{-5}$ eV$^2$/MeV. Thus, for $\Delta m^2_\odot$ in the favored region, around $8 \times 10^{-5}$ eV$^2$, and $E$ a typical $^8$B neutrino energy ($\sim$6-7 MeV), $\mathcal{H}_M$ dominates over $\mathcal{H}_V$. This means that, in first approximation, $\mathcal{H}(r \simeq 0)$ is diagonal. Thus, a $^8$B neutrino is born not only in a $\nu_e$ flavor eigenstate, but also, again in first approximation, in an eigenstate of the Hamiltonian $\mathcal{H}(r \simeq 0)$. Since $V > 0$, the neutrino will be in the heavier of the two eigenstates. Now, under the conditions where the LMA-MSW effect occurs, the propagation of a neutrino from $r \simeq 0$ to the outer edge of the Sun is adiabatic. That is, $N_e(r)$ changes sufficiently slowly that we may solve Schrödinger's equation for one $r$ at a time, and then patch together the solutions. This means that our neutrino propagates outward through the Sun as one of the $r$-dependent eigenstates of the $r$-dependent $\mathcal{H}(r)$. Since the eigenvalues of $\mathcal{H}(r)$ do not cross at any $r$, and our neutrino is born in the heavier of the two $r = 0$ eigenstates, it emerges from the Sun in the heavier of the two $\mathcal{H}_V$ eigenstates [12]. The latter is the mass eigenstate we have called $\nu_2$, given according to Eq. (1.18) by

$$\nu_2 = \nu_e \sin\theta_\odot + \nu_x \cos\theta_\odot \ . \tag{1.23}$$

Since this is an eigenstate of the vacuum Hamiltonian, the neutrino remains in it all the way to the surface of the Earth. The probability of observing the neutrino as a $\nu_e$ on Earth is then just the probability that $\nu_2$ is a $\nu_e$. That is [cf. Eq. (1.23)] [13],

$$P(\nu_e \to \nu_e) = \sin^2\theta_\odot \ . \tag{1.24}$$

We note that for $\theta_\odot < \pi/4$, this $\nu_e$ survival probability is less than $1/2$. In contrast, when matter effects are negligible, the energy-averaged survival probability in two-neutrino oscillation cannot be less than $1/2$ for any mixing angle [see Eq. (1.20)] [14].

**II. The evidence for flavor metamorphosis, and what it has taught us:** The persuasiveness of the evidence that neutrinos actually do change flavor in nature is summarized in Table 1.1. We discuss the different pieces of evidence, and what, together, they imply.

**Table 1.1:** The persuasiveness of the evidence for neutrino flavor change. The symbol $L$ denotes the distance travelled by the neutrinos. LSND is the Liquid Scintillator Neutrino Detector experiment, and MiniBooNE is an experiment designed to confirm or refute LSND.

| Neutrinos | Evidence for Flavor Change |
|---|---|
| Atmospheric | Compelling |
| Accelerator ($L = 250$ and $735\,\mathrm{km}$) | Compelling |
| Solar | Compelling |
| Reactor ($L \sim 180\,\mathrm{km}$) | Compelling |
| From Stopped $\mu^+$ Decay (LSND) | Unconfirmed by MiniBooNE |

The atmospheric neutrinos are produced in the Earth's atmosphere by cosmic rays, and then detected in an underground detector. The flux of cosmic rays that lead to neutrinos with energies above a few GeV is isotropic, so that these neutrinos are produced at the same rate all around the Earth. This can easily be shown to imply that at any underground site, the downward- and upward-going fluxes of multi-GeV neutrinos of a given flavor must be equal. That is, unless some mechanism changes the flux of neutrinos of the given flavor as they propagate, the flux coming down from zenith angle $\theta_Z$ must equal that coming up from angle $\pi - \theta_Z$ [15].

The underground Super-Kamiokande (SK) detector finds that for multi-GeV atmospheric muon neutrinos, the $\theta_Z$ event distribution looks nothing like the expected $\theta_Z \Leftrightarrow \pi - \theta_Z$ symmetric distribution. For $\cos \theta_Z \gtrsim 0.3$, the observed $\nu_\mu$ flux coming up from zenith angle $\pi - \theta_Z$ is only about half that coming down from angle $\theta_Z$ [16]. Thus, some mechanism does change the $\nu_\mu$ flux as the neutrinos travel to the detector. Since the upward-going muon neutrinos come from the atmosphere on the opposite side of the Earth from the detector, they travel much farther than the downward-going ones to reach the detector. Thus, if the muon neutrinos are oscillating away into another flavor, the upward-going ones have more distance (hence more time) in which to do so, which would explain why Flux Up < Flux Down.

If atmospheric muon neutrinos are disappearing via oscillation into another flavor, then a significant fraction of accelerator-generated muon neutrinos should disappear on their way to a sufficiently distant detector. This disappearance has been observed by both the K2K [17] and MINOS [18] experiments. Each of these experiments measures its $\nu_\mu$ flux in a detector near the neutrino source, before any oscillation is expected, and then measures it again in a detector $250\,\mathrm{km}$ from the source in the case of K2K, and $735\,\mathrm{km}$ from it in the case of MINOS. In its far detector, MINOS has observed 215 $\nu_\mu$ events in a data sample where $336 \pm 14.4$ events would have been expected, in the absence of oscillation, on the basis of the near-detector measurements. Both K2K and MINOS also find that the energy spectrum of surviving muon neutrinos in the far detector is distorted in a way that is consistent with two-neutrino oscillation.

The null results of short-baseline reactor neutrino experiments [8] imply limits on $P(\overline{\nu_e} \to \overline{\nu_\mu})$, which, assuming $CPT$ invariance, are also limits on $P(\nu_\mu \to \nu_e)$. From the latter, we know that the neutrinos into which the atmospheric, K2K, and MINOS muon neutrinos oscillate are not electron neutrinos, except possibly a small fraction of the time. All of the voluminous SK atmospheric neutrino data, corroborating data from other atmospheric neutrino experiments [19,20], K2K accelerator neutrino data, and existing MINOS accelerator neutrino data, are very well described by pure $\nu_\mu \to \nu_\tau$ quasi-two-neutrino oscillation. The allowed region for the oscillation parameters, $\Delta m^2_{\text{atm}}$ and $\sin^2 2\theta_{\text{atm}}$, which may be identified respectively with the parameters $\Delta M^2$ and $4T_\mu(1 - T_\mu)$ in Eq. (1.15), is shown in Fig. 1.1. We note that this figure implies that at least one mass eigenstate $\nu_i$ must have a mass exceeding 40 meV.

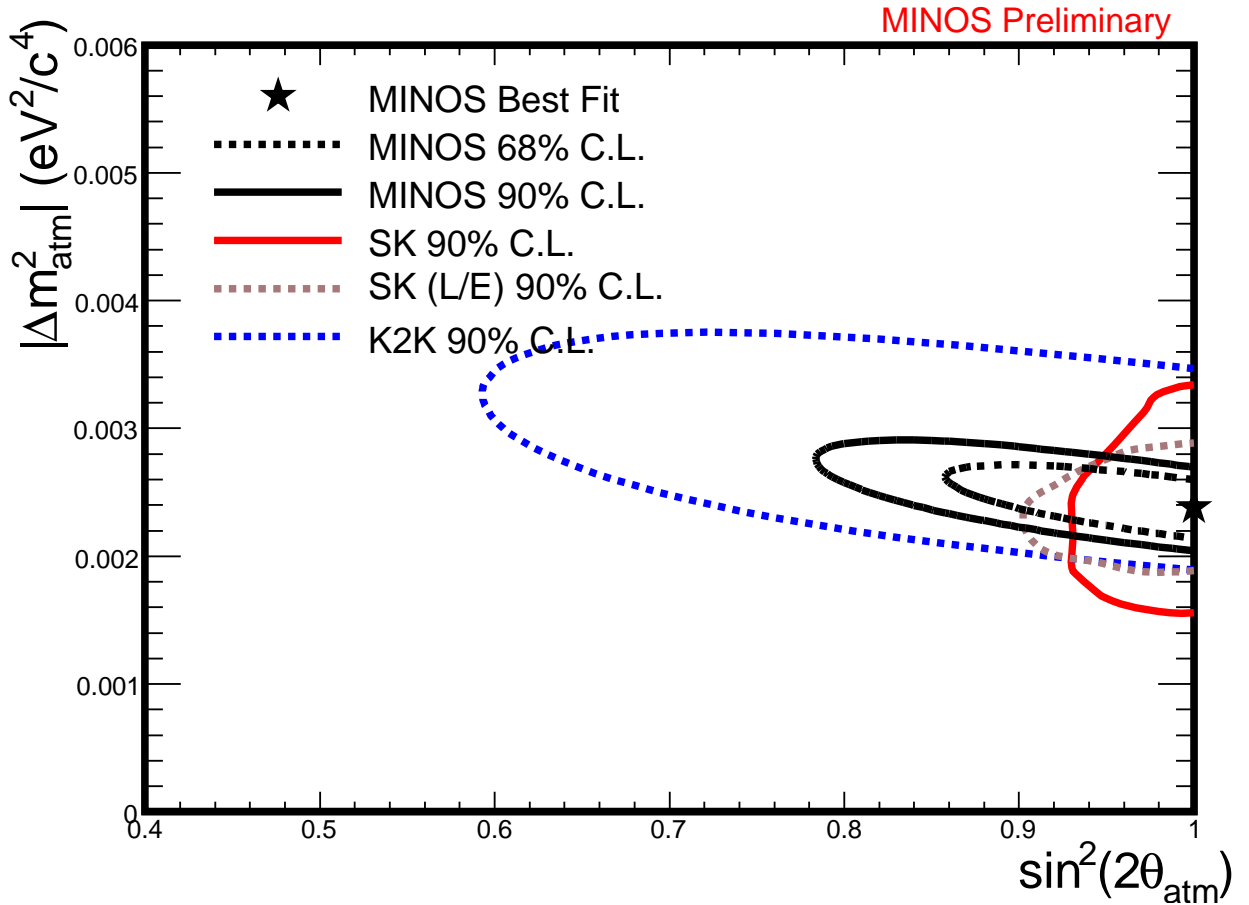

**Figure 1.1:** The region of the atmospheric oscillation parameters $\Delta m^2_{\text{atm}}$ and $\sin^2 2\theta_{\text{atm}}$ allowed by the SK, K2K, and MINOS data. The results of two different analyses of the SK ("Super K") data are shown [21].

The neutrinos created in the Sun have been detected on Earth by several experiments, as discussed by K. Nakamura in this *Review*. The nuclear processes that power the Sun make only $\nu_e$, not $\nu_\mu$ or $\nu_\tau$. For years, solar neutrino experiments had been finding that the solar $\nu_e$ flux arriving at the Earth is below the one expected from neutrino production calculations. Now, thanks especially to the Sudbury Neutrino Observatory (SNO), we have compelling evidence that the missing $\nu_e$ have simply changed into neutrinos of other flavors.

SNO has studied the flux of high-energy solar neutrinos from $^8$B decay. This experiment detects these neutrinos via the reactions

$$\nu + d \to e^- + p + p \ , \tag{1.25}$$

$$\nu + d \to \nu + p + n \ , \tag{1.26}$$

and

$$\nu + e^- \to \nu + e^- \ . \tag{1.27}$$

The first of these reactions, charged-current deuteron breakup, can be initiated only by a $\nu_e$. Thus, it measures the flux $\phi(\nu_e)$ of $\nu_e$ from $^8$B decay in the Sun. The second reaction, neutral-current deuteron breakup, can be initiated with equal cross sections by neutrinos of all active flavors. Thus, it measures $\phi(\nu_e) + \phi(\nu_{\mu,\tau})$, where $\phi(\nu_{\mu,\tau})$ is the flux of $\nu_\mu$ and/or $\nu_\tau$ from the Sun. Finally, the third reaction, neutrino electron elastic scattering, can be triggered by a neutrino of any active flavor, but $\sigma(\nu_{\mu,\tau}\, e \to \nu_{\mu,\tau}\, e) \simeq \sigma(\nu_e\, e \to \nu_e\, e)/6.5$. Thus, this reaction measures $\phi(\nu_e) + \phi(\nu_{\mu,\tau})/6.5$.

SNO finds from its observed rates for the two deuteron breakup reactions that [22]

$$\frac{\phi\,(\nu_e)}{\phi\,(\nu_e) + \phi\,(\nu_{\mu,\tau})} = 0.340 \pm 0.023\,(\text{stat})\,{}^{+0.029}_{-0.031}\,(\text{syst}) \ . \tag{1.28}$$

Clearly, $\phi(\nu_{\mu,\tau})$ is not zero. This non-vanishing $\nu_{\mu,\tau}$ flux from the Sun is "smoking-gun" evidence that some of the $\nu_e$ produced in the solar core do indeed change flavor.

Corroborating information comes from the detection reaction $\nu e^- \to \nu e^-$, studied by both SNO and SK [23].

Change of neutrino flavor, whether in matter or vacuum, does not change the total neutrino flux. Thus, unless some of the solar $\nu_e$ are changing into sterile neutrinos, the total active high-energy flux measured by the neutral-current reaction (1.26) should agree with the predicted total $^8$B solar neutrino flux based on calculations of neutrino production in the Sun. This predicted total is $(5.49^{+0.95}_{-0.81}) \times 10^6$ cm$^{-2}$s$^{-1}$ or $(4.34^{+0.71}_{-0.61}) \times 10^6$ cm$^{-2}$s$^{-1}$, depending on assumptions about the solar heavy element abundances [24]. By comparison, the total active flux measured by reaction (1.26) is $[4.94 \pm 0.21\,(\text{stat})\,{}^{+0.38}_{-0.34}\,(\text{syst})] \times 10^6$ cm$^{-2}$s$^{-1}$, in good agreement. This agreement provides evidence that neutrino production in the Sun is correctly understood, further strengthens the evidence that neutrinos really do change flavor, and strengthens the evidence that the previously-reported deficits of solar $\nu_e$ flux are due to this change of flavor.

The strongly favored explanation of $^8$B solar neutrino flavor change is the LMA-MSW effect. As pointed out after Eq. (1.24), a $\nu_e$ survival probability below 1/2, which is indicated by Eq. (1.28), requires that solar matter effects play a significant role [25]. However, from Eq. (1.22) we see that as the energy $E$ of a solar neutrino decreases, the vacuum (1st) term in the Hamiltonian $\mathcal{H}$ dominates more and more over the matter term. When we go from the $^8$B neutrinos with typical energies of $\sim$6-7 MeV to the monoenergetic $^7$Be neutrinos with energy 0.862 MeV, the matter term becomes fairly insignificant, and the $\nu_e$ survival probability is expected to be given by the vacuum oscillation formula of Eq. (1.20). In this formula, $\theta$ is to be taken as the vacuum solar mixing angle $\theta_\odot \simeq 35°$ implied by the $^8$B solar neutrino data via Eqs. (1.28) and (1.24). When averaged over the energy-line shape, the oscillatory factor $\sin^2[1.27 \, \Delta m^2(L/E)]$ is 1/2, so that from Eq. (1.20) we expect that for the $^7$Be neutrinos, P $(\nu_e \to \nu_e) \approx 0.6$.

The Borexino experiment has now provided the first real time detection of the 0.862 MeV $^7$Be solar neutrinos [26]. Borexino uses a liquid scintillator detector that detects these neutrinos via elastic neutrino-electron scattering. The experiment reports a $^7$Be $\nu_e$ counting rate of $[47 \pm 7\,(\text{stat}) \pm 12(\text{syst})]$ counts/day/100 tons. Without any flavor change, this rate would have been expected to be $[75 \pm 4]$ counts/day/100 tons. With the degree of flavor change predicted by our understanding of the $^8$B data [27] (see rough argument above), the rate would have been expected to be $[49 \pm 4]$ counts/day/100 tons. The Borexino data are in nice agreement with the latter expectation, and the Borexino Collaboration is vigorously engaged in reducing its uncertainties.

The LMA-MSW interpretation of $^8$B solar neutrino behavior implies that a substantial fraction of reactor $\overline{\nu}_e$ that travel more than a hundred kilometers should disappear into antineutrinos of other flavors. The KamLAND experiment [28], which studies reactor $\overline{\nu}_e$ that typically travel $\sim 180$ km to reach the detector, confirms this disappearance. In addition, KamLAND finds that the spectrum of the surviving $\overline{\nu}_e$ that do reach the detector is distorted, relative to the no-oscillation spectrum. As Fig. 1.2 shows, the survival probability P $(\overline{\nu}_e \to \overline{\nu}_e)$ measured by KamLAND is very well described by the hypothesis of neutrino oscillation. In particular, the measured survival probability displays the signature oscillatory behavior of the two-neutrino expression of Eq. (1.20). Ideally, the data in Fig. 1.2 would be plotted vs. $L/E$. However, KamLAND detects the $\overline{\nu}_e$ from a number of power reactors, at a variety of distances from the detector, so the distance $L$ travelled by any given $\overline{\nu}_e$ is unknown. Consequently, Fig. 1.2 plots the data vs. $L_0/E$, where $L_0 = 180$ km is a flux-weighted average travel distance. The oscillation curve and histogram in the figure take the actual distances to the individual reactors into account. Nevertheless, almost two cycles of the sinusoidal structure expected from neutrino oscillation are still plainly visible.

The region allowed by solar neutrino experiments for the two-neutrino vacuum oscillation parameters $\Delta m_\odot^2$ and $\theta_\odot$, and that allowed by KamLAND for what we believe to be the same parameters, are shown in Fig. 1.3. From this figure, we see that there is a region of overlap. This is strong evidence that the behavior of both solar neutrinos and reactor antineutrinos has been correctly understood. A joint analysis of KamLAND and solar neutrino data assuming $CPT$ invariance yields $\Delta m_\odot^2 = (7.59 \pm 0.21) \times 10^{-5}$ eV$^2$ and $\tan^2 \theta_\odot = 0.47^{+0.06}_{-0.05}$ [28].

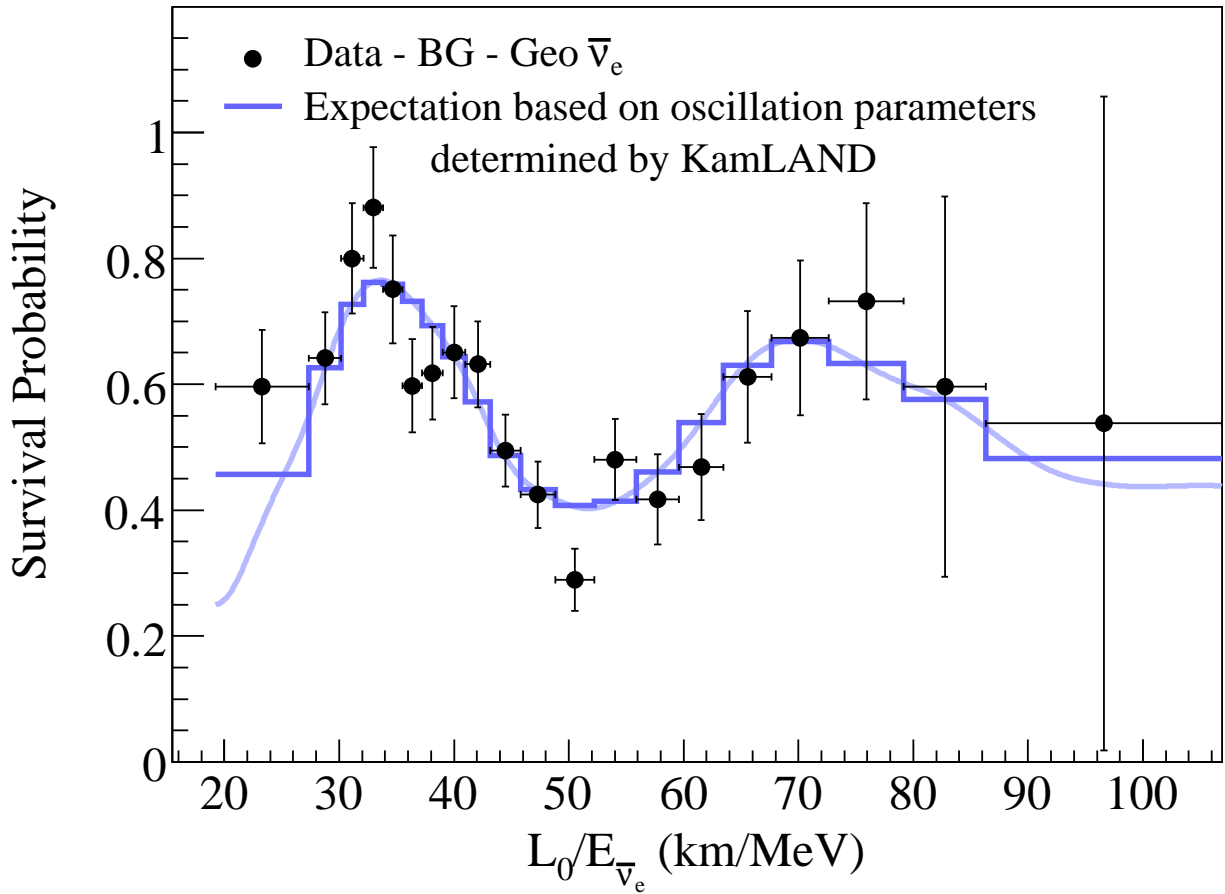

**Figure 1.2:** Ratio of the background- and geo-neutrino subtracted $\overline{\nu}_e$ spectrum to the no-oscillation expectation as a function of $L_0/E$ [28]. See text for explanation.

That $\theta_{\text{atm}}$ and $\theta_\odot$ are both large, in striking contrast to all quark mixing angles, is very interesting.

The neutrinos studied by the LSND experiment [29] come from the decay $\mu^+ \rightarrow e^+ \nu_e \overline{\nu}_\mu$ of muons at rest. While this decay does not produce $\overline{\nu}_e$, an excess of $\overline{\nu}_e$ over expected background is reported by the experiment. This excess is interpreted as due to oscillation of some of the $\overline{\nu}_\mu$ produced by $\mu^+$ decay into $\overline{\nu}_e$. The related Karlsruhe Rutherford Medium Energy Neutrino (KARMEN) experiment [30] sees no indication for such an oscillation. However, the LSND and KARMEN experiments are not identical; at LSND the neutrino travels a distance $L \approx 30$ m before detection, while at KARMEN it travels $L \approx 18$ m. The KARMEN results exclude a portion of the neutrino parameter region favored by LSND, but not all of it. A joint analysis [31] of the results of both experiments finds that a splitting $0.2 \lesssim \Delta m^2_{\text{LSND}} \lesssim 1\,\text{eV}^2$ and mixing $0.003 \lesssim \sin^2 2\theta_{\text{LSND}} \lesssim 0.03$, or a splitting $\Delta m^2_{\rm LSND} \simeq 7\,{\rm eV}^2$ and mixing $\sin^2 2\theta_{\rm LSND} \simeq 0.004$, might explain both experiments.

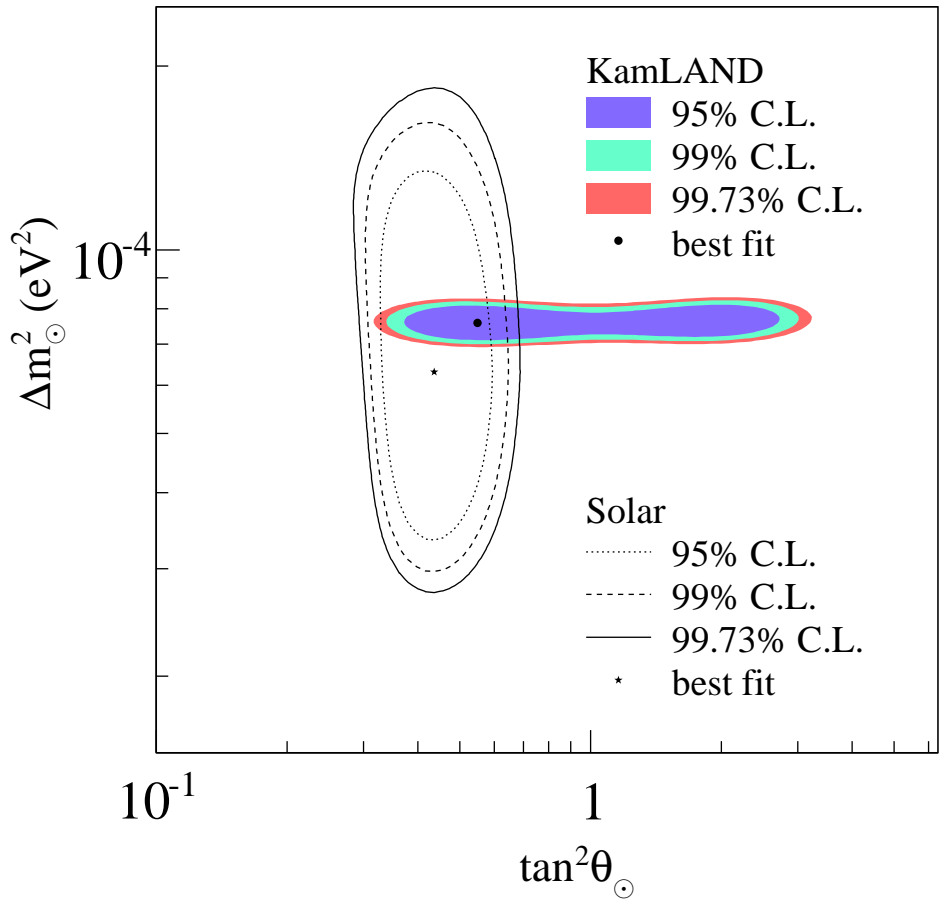

**Figure 1.3:** Regions allowed by the "solar" neutrino oscillation parameters by KamLAND and by solar neutrino experiments [28].

To confirm or exclude the LSND oscillation signal, the MiniBooNE experiment was launched. MiniBooNE studies $\nu_\mu$ and $\overline{\nu_\mu}$ that travel a distance $L$ of 540 m and have a typical energy $E$ of 700 MeV, so that $L/E$ is of order 1 km/GeV as in LSND. MiniBooNE's first results [32], regarding a search for $\nu_\mu \to \nu_e$ oscillation in a $\nu_\mu$ beam, do not confirm LSND. For neutrino energies $475 < E < 3000\,{\rm MeV}$, there is no significant excess of events above background. A joint analysis of the MiniBooNE data at these energies and the LSND data excludes at 98% CL two-neutrino $\overline{\nu_\mu} \to \overline{\nu_e}$ oscillation as an explanation of the LSND $\overline{\nu_e}$ excess. To be sure, there is an excess of MiniBooNE $\nu_e$ candidate events below 475 MeV. This low-energy excess cannot be explained by two-neutrino oscillation, and its source is being studied. Possibilities include an unidentified background, a Standard

Model effect that has been proposed only recently [33], and many-neutrino oscillation with a *CP* violation that allows the *antineutrino* oscillation reported by LSND to differ from the *neutrino* results reported so far by MiniBooNE [34].

The MiniBooNE detector is illuminated by both the neutrino beam constructed for the purpose, and the beam that is aimed at the MINOS detector. The distance $L$ to MiniBooNE from the neutrino source is 40% larger in the latter beam than in the former. When matter effects may be neglected, the probability of oscillation depends on $L$ and the beam energy $E$ only through $L/E$ [cf. Eq. (1.9)]. Thus, if the low-energy excess seen by MiniBooNE is neutrino oscillation, it should appear at a 40% higher energy in the beam directed at MINOS than in MiniBooNE's own beam. Whether it does or not is under investigation.

The regions of neutrino parameter space favored or excluded by various neutrino oscillation experiments are shown in Fig. 1.4.

**III. Neutrino spectra and mixings:**  If there are only three neutrino mass eigenstates, $\nu_1, \nu_2$, and $\nu_3$, then there are only three mass splittings $\Delta m_{ij}^2$, and they obviously satisfy

$$\Delta m_{32}^2 + \Delta m_{21}^2 + \Delta m_{13}^2 = 0 \ . \tag{1.29}$$

However, as we have seen, the $\Delta m^2$ values required to explain the flavor changes of the atmospheric, solar, and LSND neutrinos are of three different orders of magnitude. Thus, they cannot possibly obey the constraint of Eq. (1.29). If all of the reported changes of flavor are genuine, then nature must contain at least four neutrino mass eigenstates [35]. As explained in Sec. I, one linear combination of these mass eigenstates would have to be sterile.

If further MiniBooNE results do not confirm the LSND oscillation, then nature may well contain only three neutrino mass eigenstates. The neutrino spectrum then contains two mass eigenstates separated by the splitting $\Delta m_\odot^2$ needed to explain the solar and KamLAND data, and a third eigenstate separated from the first two by the larger splitting $\Delta m_{\rm atm}^2$ called for by the atmospheric, MINOS, and K2K data. Current experiments do not tell us whether the solar pair — the two eigenstates separated by $\Delta m_\odot^2$ — is at the bottom or the top of the spectrum. These two possibilities are usually referred to, respectively, as a normal and an inverted spectrum. The study of flavor changes of accelerator-generated neutrinos and antineutrinos that pass through matter can discriminate between these two spectra (see Sec. V). If the solar pair is at the bottom, then the spectrum is of the form shown in Fig. 1.5. There we include the approximate flavor content of each mass eigenstate, the flavor-$\alpha$ fraction of eigenstate $\nu_i$ being simply $|\langle \nu_\alpha | \nu_i \rangle|^2 = |U_{\alpha i}|^2$. The flavor content shown assumes that the atmospheric mixing angle is maximal, which gives the best fit to the atmospheric data [16] and, as indicated in Fig. 1.1, to the MINOS data. The content shown also takes into account the now-established LMA-MSW explanation of solar neutrino behavior. For simplicity, it neglects the small, as-yet-unknown $\nu_e$ fraction of $\nu_3$ (see below).

When there are only three neutrino mass eigenstates, and the corresponding three familiar neutrinos of definite flavor, the leptonic mixing matrix $U$ can be written as

$$
U = \begin{array}{c} \nu_e \\ \nu_\mu \\ \nu_\tau \end{array}
\begin{array}{ccc} \nu_1 & \nu_2 & \nu_3 \end{array}
\left[ \begin{array}{ccc}
c_{12}c_{13} & s_{12}c_{13} & s_{13}e^{-i\delta} \\
-s_{12}c_{23}-c_{12}s_{23}s_{13}e^{i\delta} & c_{12}c_{23}-s_{12}s_{23}s_{13}e^{i\delta} & s_{23}c_{13} \\
s_{12}s_{23}-c_{12}c_{23}s_{13}e^{i\delta} & -c_{12}s_{23}-s_{12}c_{23}s_{13}e^{i\delta} & c_{23}c_{13}
\end{array} \right]
$$
$$
\times \operatorname{diag}\left( e^{i\alpha_1/2},\ e^{i\alpha_2/2},\ 1 \right) \ . \tag{1.30}
$$

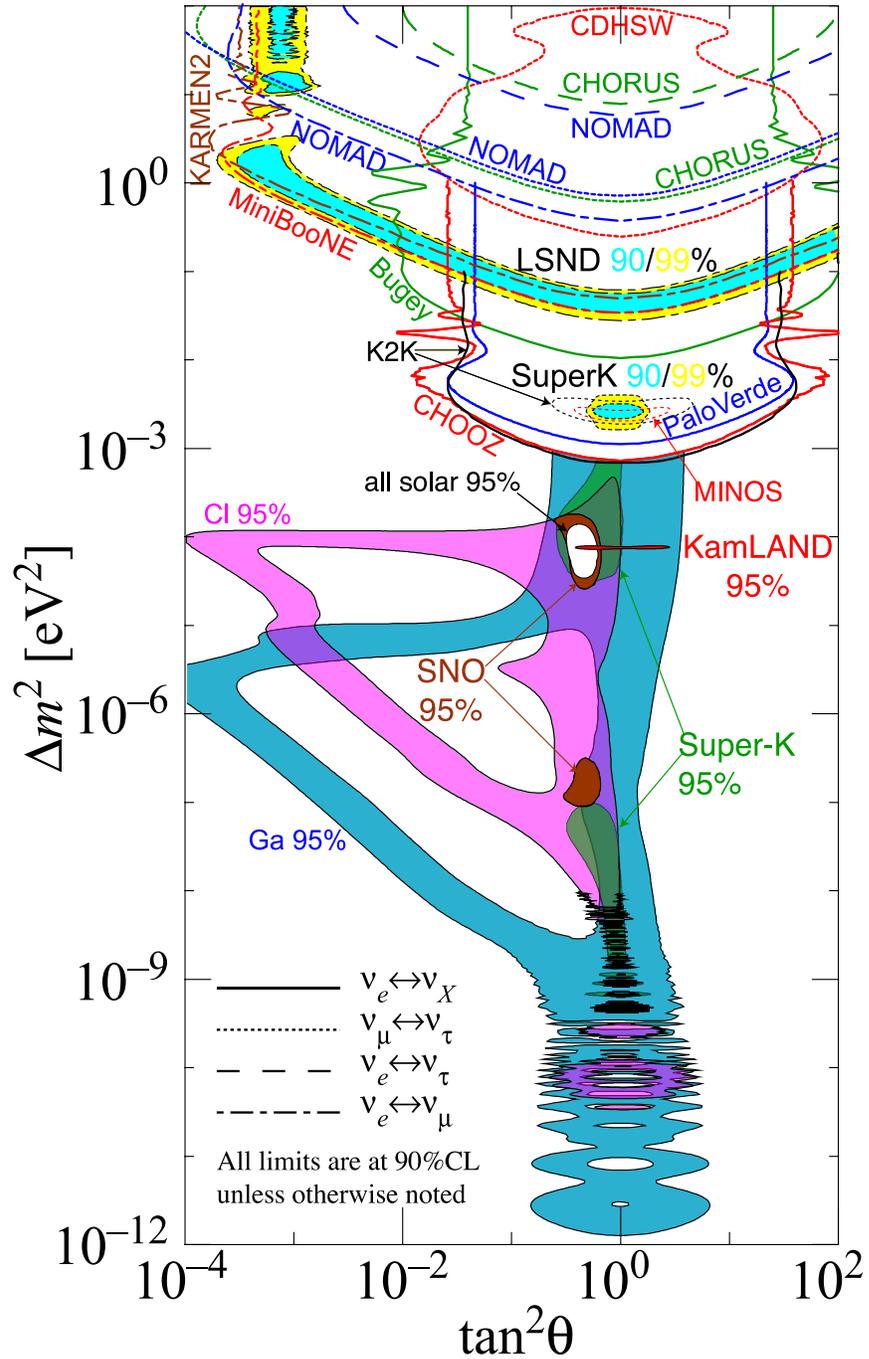

**Figure 1.4:** The regions of squared-mass splitting and mixing angle favored or excluded by various experiments. This figure was contributed by H. Murayama (University of California, Berkeley). References to the data used in the figure can be found at http://hitoshi.berkeley.edu/neutrino/.

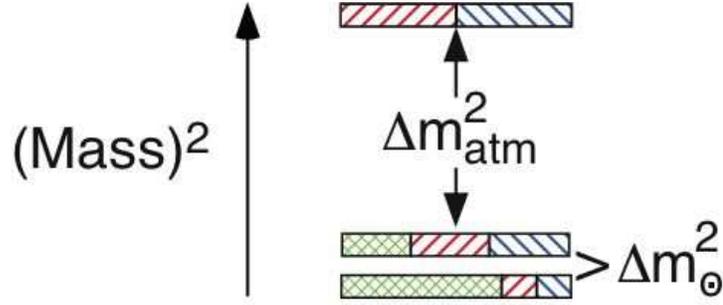

**Figure 1.5:** A three-neutrino squared-mass spectrum that accounts for the observed flavor changes of solar, reactor, atmospheric, and long-baseline accelerator neutrinos. The $\nu_e$ fraction of each mass eigenstate is crosshatched, the $\nu_\mu$ fraction is indicated by right-leaning hatching, and the $\nu_\tau$ fraction by left-leaning hatching.

Here, $\nu_1$ and $\nu_2$ are the members of the solar pair, with $m_2 > m_1$, and $\nu_3$ is the isolated neutrino, which may be heavier or lighter than the solar pair. Inside the matrix, $c_{ij} \equiv \cos\theta_{ij}$ and $s_{ij} \equiv \sin\theta_{ij}$, where the three $\theta_{ij}$'s are mixing angles. The quantities $\delta$, $\alpha_1$, and $\alpha_2$ are $CP$-violating phases. The phases $\alpha_1$ and $\alpha_2$, known as Majorana phases, have physical consequences only if neutrinos are Majorana particles, identical to their antiparticles. Then these phases influence neutrinoless double-beta decay [see Sec. IV] and other processes [36]. However, as we see from Eq. (1.9), $\alpha_1$ and $\alpha_2$ do not affect neutrino oscillation, regardless of whether neutrinos are Majorana particles. Apart from the phases $\alpha_1$, $\alpha_2$, which have no quark analogues, the parametrization of the leptonic mixing matrix in Eq. (1.30) is identical to that [37] advocated for the quark mixing matrix by Ceccucci, Ligeti, and Sakai in their article in this *Review*.

From bounds on the short-distance oscillation of reactor $\overline{\nu}_e$ [8] and other data, at $2\,\sigma$, $|U_{e3}|^2 \lesssim 0.032$ [38]. (Thus, the $\nu_e$ fraction of $\nu_3$ would have been too small to see in Fig. 1.5; this is the reason it was neglected.) From Eq. (1.30), we see that the bound on $|U_{e3}|^2$ implies that $s_{13}^2 \lesssim 0.032$. From Eq. (1.30), we also see that the $CP$-violating phase $\delta$, which is the sole phase in the $U$ matrix that can produce $CP$ violation in neutrino oscillation, enters $U$ only in combination with $s_{13}$. Thus, the size of $CP$ violation in oscillation will depend on $s_{13}$.

Given that $s_{13}$ is small, Eqs. (1.30), (1.15), and (1.17) imply that the atmospheric mixing angle $\theta_{\mathrm{atm}}$ extracted from $\nu_\mu$ disappearance measurements is approximately $\theta_{23}$, while Eqs. (1.30) and (1.18) (with $\nu_\alpha = \nu_e$ and $\theta = \theta_\odot$) imply that $\theta_\odot \simeq \theta_{12}$.

***IV. The neutrino-antineutrino relation:***   Unlike quarks and charged leptons, neutrinos may be their own antiparticles. Whether they are depends on the nature of the physics that gives them mass.

In the Standard Model (SM), neutrinos are assumed to be massless. Now that we know they do have masses, it is straightforward to extend the SM to accommodate these masses in the same way that this model accommodates quark and charged lepton masses. When a neutrino $\nu$ is assumed to be massless, the SM does not contain the chirally right-handed neutrino field $\nu_R$, but only the left-handed field $\nu_L$ that couples to the $W$ and $Z$ bosons. To accommodate the $\nu$ mass in the same manner as quark masses are accommodated, we add $\nu_R$ to the Model. Then we may construct the "Dirac mass term"

$$\mathcal{L}_D = -m_D \,\overline{\nu}_L \,\nu_R + h.c. \quad , \qquad (1.31)$$

in which $m_D$ is a constant. This term, which mimics the mass terms of quarks and charged leptons, conserves the lepton number $L$ that distinguishes neutrinos and negatively-charged leptons on the one hand from antineutrinos and positively-charged leptons on the other. Since everything else in the SM also conserves $L$, we then have an $L$-conserving world. In such a world, each neutrino mass eigenstate $\nu_i$ differs from its antiparticle $\overline{\nu}_i$, the difference being that $L(\overline{\nu}_i) = -L(\nu_i)$. When $\overline{\nu}_i \neq \nu_i$, we refer to the $\nu_i - \overline{\nu}_i$ complex as a "Dirac neutrino."

Once $\nu_R$ has been added to our description of neutrinos, a "Majorana mass term,"

$$\mathcal{L}_M = -m_R \,\overline{\nu_R^c} \,\nu_R + h.c. \quad , \qquad (1.32)$$

can be constructed out of $\nu_R$ and its charge conjugate, $\nu_R^c$. In this term, $m_R$ is another constant. Since both $\nu_R$ and $\overline{\nu_R^c}$ absorb $\nu$ and create $\overline{\nu}$, $\mathcal{L}_M$ mixes $\nu$ and $\overline{\nu}$. Thus, a Majorana mass term does not conserve $L$. In somewhat the same way that, neglecting $CP$ violation, $K^0 - \overline{K^0}$ mixing causes the neutral kaon mass eigenstates to be the self-conjugate states $(K^0 \pm \overline{K^0})/\sqrt{2}$, the $\nu - \bar{\nu}$ mixing induced by a Majorana mass term causes the neutrino mass eigenstates to be self-conjugate: $\overline{\nu}_i = \nu_i$. That is, for a given helicity $h$, $\overline{\nu}_i(h) = \nu_i(h)$. We then refer to $\nu_i$ as a "Majorana neutrino."

Suppose the right-handed neutrinos required by Dirac mass terms have been added to the SM. If we insist that this extended SM conserve $L$, then, of course, Majorana mass terms are forbidden. However, if we do not impose $L$ conservation, but require only the general principles of gauge invariance and renormalizability, then Majorana mass terms like that of Eq. (1.32) are expected to be present. As a result, $L$ is violated, and neutrinos are Majorana particles [39].

In the see-saw mechanism [40], which is the most popular explanation of why neutrinos — although massive — are nevertheless so light, both Dirac and Majorana mass terms are present. Hence, the neutrinos are Majorana particles. However, while half of them are the familiar light neutrinos, the other half are extremely heavy Majorana particles referred to as the $N_i$, with masses possibly as large as the GUT scale. The $N_i$ may have played a crucial role in baryogenesis in the early universe, as we shall discuss in Sec. V.

How can the theoretical expectation that nature contains Majorana mass terms, so that $L$ is violated and neutrinos are Majorana particles, be confirmed experimentally? The promising approach is to search for neutrinoless double-beta decay ($0\nu\beta\beta$). This is the process $(A, Z) \to (A, Z + 2) + 2e^-$, in which a nucleus containing $A$ nucleons, $Z$ of which are protons, decays to a nucleus containing $Z + 2$ protons by emitting two electrons. While $0\nu\beta\beta$ can in principle receive contributions from a variety of mechanisms (R-parity-violating supersymmetric couplings, for example), it is easy to show explicitly that its observation at any non-vanishing rate would imply that nature contains at least one Majorana neutrino mass term [41]. The neutrino mass eigenstates must then be Majorana neutrinos.

Quarks and charged leptons cannot have Majorana mass terms, because such terms mix fermion and antifermion, and $q \leftrightarrow \overline{q}$ or $\ell \leftrightarrow \overline{\ell}$ would not conserve electric charge. Thus, the discovery of $0\nu\beta\beta$ would demonstrate that the physics of neutrino masses is unlike that of the masses of all other fermions.

The dominant mechanism for $0\nu\beta\beta$ is expected to be the one depicted in Fig. 1.6. There, a pair of virtual $W$ bosons are emitted by the parent nucleus, and then these $W$ bosons exchange one or another of the light neutrino mass eigenstates $\nu_i$ to produce the outgoing electrons. The $0\nu\beta\beta$ amplitude is then a sum over the contributions of the different $\nu_i$. It is assumed that the interactions at the two leptonic $W$ vertices are those of the SM.

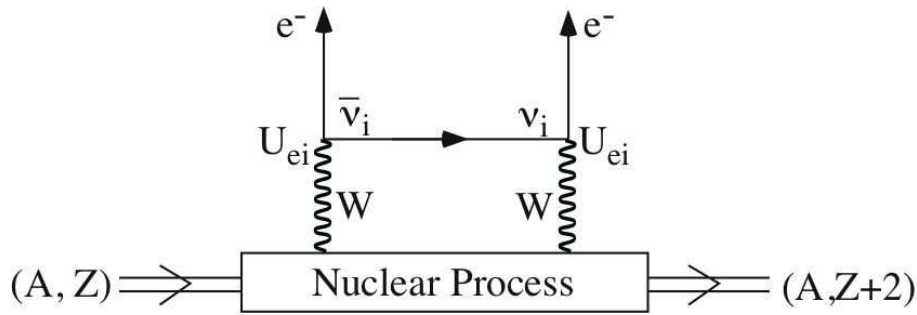

**Figure 1.6:** The dominant mechanism for $0\nu\beta\beta$. The diagram does not exist unless $\overline{\nu}_i = \nu_i$.

Since the exchanged $\nu_i$ is created together with an $e^-$, the left-handed SM current that creates it gives it the helicity we associate, in common parlance, with an "antineutrino." That is, the $\nu_i$ is almost totally right-handed, but has a small left-handed-helicity component, whose amplitude is of order $m_i/E$, where $E$ is the $\nu_i$ energy. At the vertex where this $\nu_i$ is absorbed, the absorbing left-handed SM current can absorb only its small left-handed-helicity component without further suppression. Consequently, the $\nu_i$-exchange contribution to the $0\nu\beta\beta$ amplitude is proportional to $m_i$. From Fig. 1.6, we see that this contribution is also proportional to $U_{ei}^2$. Thus, summing over the contributions of all the $\nu_i$, we conclude that the amplitude for $0\nu\beta\beta$ is proportional to the quantity

$$\left| \sum_i m_i\, U_{ei}^2 \right| \equiv |<m_{\beta\beta}>| \quad, \tag{1.33}$$

commonly referred to as the "effective Majorana mass for neutrinoless double-beta decay" [42].

To how small an $|<m_{\beta\beta}>|$ should a $0\nu\beta\beta$ search be sensitive? In answering this question, it makes sense to assume there are only three neutrino mass eigenstates — if there are more, $|<m_{\beta\beta}>|$ might be larger. Suppose that there are just three mass eigenstates, and that the solar pair, $\nu_1$ and $\nu_2$, is at the top of the spectrum, so that we have an inverted spectrum. If the various $\nu_i$ are not much heavier than demanded by the observed splittings $\Delta m_{\rm atm}^2$ and $\Delta m_\odot^2$, then in $|<m_{\beta\beta}>|$, Eq. (1.33), the contribution of $\nu_3$ may be neglected, because both $m_3$ and $|U_{e3}^2| = s_{13}^2$ are small. From Eqs. (1.33) and (1.30), approximating $c_{13}$ by unity, we then have that

$$|<m_{\beta\beta}>| \simeq m_0 \sqrt{1 - \sin^2 2\theta_\odot \sin^2\left(\frac{\Delta\alpha}{2}\right)} \quad. \tag{1.34}$$

Here, $m_0$ is the average mass of the members of the solar pair, whose splitting will be invisible in a practical $0\nu\beta\beta$ experiment, and $\Delta\alpha \equiv \alpha_2 - \alpha_1$ is a $CP$-violating phase. Although $\Delta\alpha$ is completely unknown, we see from Eq. (1.34) that

$$|<m_{\beta\beta}>| \geq m_0 \cos 2\theta_\odot \quad. \tag{1.35}$$

Now, in an inverted spectrum, $m_0 \geq \sqrt{\Delta m_{\rm atm}^2}$. At 90% CL, $\sqrt{\Delta m_{\rm atm}^2} > 45\,{\rm meV}$ [see Fig. 1.1], while at 95% CL, $\cos 2\theta_\odot > 0.25$ [see Fig. 1.3]. Thus, if neutrinos are Majorana particles, and the spectrum is as we have assumed, a $0\nu\beta\beta$ experiment sensitive to $|<m_{\beta\beta}>| \gtrsim 10$ meV would have an excellent chance of observing a signal. If the spectrum is inverted, but the $\nu_i$ masses are larger than the $\Delta m_{\rm atm}^2$- and $\Delta m_\odot^2$-demanded minimum values we have assumed above, then once again $|<m_{\beta\beta}>|$ is larger than 10 meV [43], and an experiment sensitive to 10 meV still has an excellent chance of seeing a signal.

If the solar pair is at the bottom of the spectrum, rather than at the top, then $|<m_{\beta\beta}>|$ is not as tightly constrained, and can be anywhere from the present bound of 0.3–1.0 eV down to invisibly small [43,44]. For a discussion of the present bounds, see the article by Vogel and Piepke in this *Review* [45].

**V. Questions to be answered:** The strong evidence for neutrino flavor metamorphosis — hence neutrino mass — opens many questions about the neutrinos. These questions, which hopefully will be answered by future experiments, include the following:

*i) How many neutrino species are there? Do sterile neutrinos exist?*

This question is being addressed by the MiniBooNE experiment [32]. If MiniBooNE's final result is positive, the implications will be far-reaching. We will have learned that either there are more than three neutrino species and at least one of these species is sterile, or else there is an even more amazing departure from what has been our picture of the neutrino world.

*ii) What are the masses of the mass eigenstates $\nu_i$?*

Assuming there are only three $\nu_i$, we need to find out whether the solar pair, $\nu_{1,2}$, is at the bottom of the spectrum or at its top. This can be done by exploiting matter effects in long-baseline neutrino and antineutrino oscillations. These matter effects will determine the sign one wishes to learn — that of $\{m_3^2 - [(m_2^2 + m_1^2)/2]\}$ — relative to a sign that is already known — that of the interaction energy of Eq. (1.21). Grand unified theories favor a spectrum with the closely spaced solar pair at the bottom [46]. The neutrino spectrum would then resemble the spectra of the quarks, to which grand unified theories relate the neutrinos. A neutrino spectrum with the closely spaced solar pair at the top would be quite un-quark-like, and would suggest the existence of a new symmetry that leads to the near degeneracy at the top of the spectrum.

While flavor-change experiments can determine a spectral pattern such as the one in Fig. 1.5, they cannot tell us the distance of the entire pattern from the zero of squared-mass. One might discover that distance via study of the $\beta$ energy spectrum in tritium $\beta$ decay, if the mass of some $\nu_i$ with appreciable coupling to an electron is large enough to be within reach of a feasible experiment. One might also gain some information on the distance from zero by measuring $|<m_{\beta\beta}>|$, the effective Majorana mass for neutrinoless double-beta decay [43–45] (see Vogel and Piepke in this *Review*). Finally, one might obtain information on this distance from cosmology or astrophysics. Indeed, from current cosmological data and some cosmological assumptions, it is already concluded that [47]

$$\sum_i m_i < (0.17 - 2.0) \text{ eV} \ .$$ (1.36)

Here, the sum runs over the masses of all the light neutrino mass eigenstates $\nu_i$ that may exist and that were in thermal equilibrium in the early universe. The range quoted in Eq. (1.36) reflects the dependence of this upper bound on the underlying cosmological assumptions and on which data are used [47].

If there are just three $\nu_i$, and their spectrum is either the one shown in Fig. 1.5 or its inverted version, then Eq. (1.36) implies that the mass of the heaviest $\nu_i$, Mass [Heaviest $\nu_i$], cannot exceed $(0.07 - 0.7)$ eV. Moreover, Mass [Heaviest $\nu_i$] obviously cannot be less than $\sqrt{\Delta m_{\text{atm}}^2}$, which in turn is not less than 0.04 eV, as previously noted. Thus, if the cosmological assumptions behind Eq. (1.36) are correct, then

$$0.04 \text{ eV} < \text{Mass [Heaviest } \nu_i] < (0.07 - 0.7) \text{ eV} \ .$$ (1.37)

*iii*) *Are the neutrino mass eigenstates Majorana particles?*

The confirmed observation of neutrinoless double-beta decay would establish that the answer is "yes." If there are only three $\nu_i$, knowledge that the spectrum is inverted and a definitive upper bound on $|<m_{\beta\beta}>|$ that is well below 0.01 eV would establish (barring exotic contributions to $0\nu\beta\beta$) that the answer is "no" [see discussion after Eq. (1.35)] [43,44].

*iv*) *What are the mixing angles in the leptonic mixing matrix $U$?*

The solar mixing angle $\theta_\odot \simeq \theta_{12}$ is already rather well determined.

The atmospheric mixing angle $\theta_{\text{atm}} \simeq \theta_{23}$ is constrained by the most stringent analysis to lie, at 90% CL, in the region where $\sin^2 2\theta_{\text{atm}} > 0.92$ [16]. This region is still fairly large: $37°$ to $53°$. A more precise value of $\sin^2 2\theta_{\text{atm}}$, and, in particular, its deviation from unity, can be sought in precision long-baseline $\nu_\mu$ disappearance experiments. If $\sin^2 2\theta_{\text{atm}} \neq 1$, so that $\theta_{\text{atm}} \neq 45°$, one can determine whether it lies below or above $45°$ with the help of a reactor $\overline{\nu}_e$ experiment [48,49]. Once we know whether the neutrino spectrum is normal or inverted, this determination will tell us whether the heaviest mass eigenstate is more $\nu_\tau$ than $\nu_\mu$, as naively expected, or more $\nu_\mu$ than $\nu_\tau$ [cf. Eq. (1.30)].

A knowledge of the small mixing angle $\theta_{13}$ is important not only to help complete our picture of leptonic mixing, but also because, as Eq. (1.30) made clear, all $CP$-violating effects of the phase $\delta$ are proportional to $\sin\theta_{13}$. Thus, a knowledge of the order of magnitude of $\theta_{13}$ would guide the planning of experiments to probe $CP$ violation. From Eq. (1.30), we recall that $\sin^2 \theta_{13}$ is the $\nu_e$ fraction of $\nu_3$. The $\nu_3$ is the isolated neutrino that lies at one end of the atmospheric squared-mass gap $\Delta m^2_{\text{atm}}$, so an experiment seeking to measure $\theta_{13}$ should have an $L/E$ that makes it sensitive to $\Delta m^2_{\text{atm}}$, and should involve $\nu_e$. Planned approaches include a sensitive search for the disappearance of reactor $\overline{\nu}_e$ while they travel a distance $L \sim 1$ km, and an accelerator neutrino search for $\nu_\mu \to \nu_e$ and $\overline{\nu_\mu} \to \overline{\nu_e}$ with a beamline $L >$ several hundred km.

If LSND is confirmed, then (barring the still more revolutionary) the matrix $U$ is at least $4 \times 4$, and contains many more than three angles. A rich program, including short baseline experiments with multiple detectors, will be needed to learn about both the squared-mass spectrum and the mixing matrix.

Given the large sizes of $\theta_{\text{atm}}$ and $\theta_\odot$, we already know that leptonic mixing is very different from its quark counterpart, where all the mixing angles are small. This difference, and the striking contrast between the tiny neutrino masses and the very much larger quark masses, suggest that the physics underlying neutrino masses and mixing may be very different from the physics behind quark masses and mixing.

*v*) *Does the behavior of neutrinos violate $CP$?*

From Eqs. (1.9), (1.13), and (1.30), we see that if the $CP$-violating phase $\delta$ and the small mixing angle $\theta_{13}$ are both non-vanishing, there will be $CP$-violating differences between neutrino and antineutrino oscillation probabilities. Observation of these differences would establish that $CP$ violation is not a peculiarity of quarks.

The $CP$-violating difference $P(\nu_\alpha \to \nu_\beta) - P(\overline{\nu}_\alpha \to \overline{\nu}_\beta)$ between "neutrino" and "antineutrino" oscillation probabilities is independent of whether the mass eigenstates $\nu_i$ are Majorana or Dirac particles. To study $\nu_\mu \to \nu_e$ with a super-intense but conventionally generated neutrino beam, for example, one would create the beam via the process $\pi^+ \to \mu^+ \nu_i$, and detect it via $\nu_i + \text{target} \to e^- + \dots$. To study $\overline{\nu}_\mu \to \overline{\nu}_e$, one would create the beam via $\pi^- \to \mu^- \overline{\nu}_i$, and detect it via $\overline{\nu}_i + \text{target} \to e^+ + \dots$. Whether $\overline{\nu}_i = \nu_i$ or not, the amplitudes for the latter two processes are proportional to $U_{\mu i}$ and $U_{ei}^*$, respectively. In contrast, the amplitudes for their $\nu_\mu \to \nu_e$ counterparts are proportional to $U_{\mu i}^*$ and $U_{ei}$. As this illustrates, Eq. (1.13) relates "neutrino" and "antineutrino" oscillation probabilities even when the neutrino mass eigenstates are their own antiparticles.

The baryon asymmetry of the universe could not have developed without some violation of $CP$ during the universe's early history. The one known source of $CP$ violation — the complex phase in the quark mixing matrix — could not have produced sufficiently large effects. Thus, perhaps *leptonic $CP$* violation is responsible for the baryon asymmetry. The see-saw mechanism predicts very heavy Majorana neutral leptons $N_i$ (see Sec. IV), which would have been produced in the Big Bang. Perhaps $CP$ violation in the leptonic decays of an $N_i$ led to the inequality

$$\Gamma\left(N_i \to \ell^+ + \dots\right) \neq \Gamma\left(N_i \to \ell^- + \dots\right) \quad , \tag{1.38}$$

which would have resulted in unequal numbers of $\ell^+$ and $\ell^-$ in the early universe [50]. This leptogenesis could have been followed by nonperturbative SM processes that would have converted the lepton asymmetry, in part, into the observed baryon asymmetry [51].

While the connection between the $CP$ violation that would have led to leptogenesis, and that which we hope to observe in neutrino oscillation, is model-dependent, it is not likely that we have either of these without the other [52], because in the see-saw picture, these two $CP$ violations both arise from the same matrix of coupling constants. This makes the search for $CP$ violation in neutrino oscillation very interesting indeed. Depending on the rough size of $\theta_{13}$, this $CP$ violation may be observable with a very intense conventional neutrino beam, or may require a "neutrino factory," whose neutrinos come from the decay of stored muons or radioactive nuclei. The detailed study of $CP$ violation may require a neutrino factory in any case.

With a conventional beam, one would seek $CP$ violation, and try to determine whether the mass spectrum is normal or inverted, by studying the oscillations $\nu_\mu \to \nu_e$ and $\overline{\nu_\mu} \to \overline{\nu_e}$. The appearance probability for $\nu_e$ in a beam that is initially $\nu_\mu$ can be written for $\sin^2 2\theta_{13} < 0.2$ [53]

$$P\left(\nu_\mu \to \nu_e\right) \cong \sin^2 2\theta_{13}\, T_1 - \alpha \sin 2\theta_{13}\, T_2 + \alpha \sin 2\theta_{13}\, T_3 + \alpha^2 T_4 \quad . \tag{1.39}$$

Here, $\alpha \equiv \Delta m_{21}^2 / \Delta m_{31}^2$ is the small ($\sim 1/30$) ratio between the solar and atmospheric squared-mass splittings, and

$$T_1 = \sin^2 \theta_{23} \frac{\sin^2\left[(1-x)\,\Delta\right]}{(1-x)^2} \quad , \tag{1.40}$$

$$T_2 = \sin \delta \sin 2\theta_{12} \sin 2\theta_{23} \sin \Delta \frac{\sin(x\Delta)}{x} \frac{\sin\left[(1-x)\,\Delta\right]}{(1-x)} \quad , \tag{1.41}$$

$$T_3 = \cos\delta \sin 2\theta_{12} \sin 2\theta_{23} \cos\Delta \frac{\sin\left(x\Delta\right)}{x} \frac{\sin\left[\left(1-x\right)\Delta\right]}{\left(1-x\right)} \quad, \tag{1.42}$$

and

$$T_4 = \cos^2\theta_{23} \sin^2 2\theta_{12} \frac{\sin^2\left(x\Delta\right)}{x^2} \quad. \tag{1.43}$$

In these expressions, $\Delta \equiv \Delta m_{31}^2 L/4E$ is the kinematical phase of the oscillation. The quantity $x \equiv 2\sqrt{2} G_F N_e E/\Delta m_{31}^2$, with $G_F$ the Fermi coupling constant and $N_e$ the electron number density, is a measure of the importance of the matter effect resulting from coherent forward-scattering of electron neutrinos from ambient electrons as the neutrinos travel through the earth from the source to the detector [cf. Sec. I]. In the appearance probability $P(\nu_\mu \to \nu_e)$, the $T_1$ term represents the oscillation due to the atmospheric-mass-splitting scale, the $T_4$ term represents the oscillation due to the solar-mass-splitting scale, and the $T_2$ and $T_3$ terms are the $CP$-violating and $CP$-conserving interference terms, respectively.

The probability for the corresponding antineutrino oscillation, $P(\overline{\nu_\mu} \to \overline{\nu_e})$, is the same as the probability $P(\nu_\mu \to \nu_e)$ given by Eqs. (1.39)–(1.43), but with the signs in front of both $x$ and $\sin\delta$ reversed: both the matter effect and $CP$ violation lead to a difference between the $\nu_\mu \to \nu_e$ and $\overline{\nu_\mu} \to \overline{\nu_e}$ oscillation probabilities. In view of the dependence of $x$ on $\Delta m_{31}^2$, and in particular on the sign of $\Delta m_{31}^2$, the matter effect can reveal whether the neutrino mass spectrum is normal or inverted. However, to determine the nature of the spectrum, and to establish the presence of $CP$ violation, it obviously will be necessary to disentangle the matter effect from $CP$ violation in the neutrino-antineutrino oscillation probability difference that is actually observed. To this end, complementary measurements will be extremely important. These can take advantage of the differing dependences on the matter effect and on $CP$ violation in $P(\nu_\mu \to \nu_e)$.

*vi) Will we encounter the completely unexpected?*

The study of neutrinos has been characterized by surprises. It would be surprising if further surprises were not in store. The possibilities include new, non-Standard-Model interactions, unexpectedly large magnetic and electric dipole moments [54], unexpectedly short lifetimes, and violations of $CPT$ invariance, Lorentz invariance, or the equivalence principle.

The questions we have discussed, and other questions about the world of neutrinos, will be the focus of a major experimental program in the years to come.

**Acknowledgements**

I am grateful to Susan Kayser for her crucial role in the production of this manuscript.